%% file: Siebel.tex
\documentclass{ws-p9-75x6-50}
\usepackage{float}

\begin{document}

\title{Characteristic numerical relativity applied to hydrodynamic
studies of neutron stars}
 
\author{F. Siebel, J.A. Font, E. M\"uller}
 
\address{Max-Planck-Institut f\"ur Astrophysik,
Karl-Schwarzschild-Str. 1,
D-85741 Garching, Germany\\
E-mail: florian@mpa-garching.mpg.de\\
font@mpa-garching.mpg.de\\
ewald@mpa-garching.mpg.de}

\author{P. Papadopoulos}

\address{School of Computer Science and Mathematics,
University of Portsmouth,        
Portsmouth, PO1 2EG, 
UK\\
E-mail: philippos.papadopoulos@port.ac.uk}  


\maketitle

\abstracts{We present tests and results of a new axisymmetric, fully
general relativistic code capable of solving the coupled
Einstein-matter system for a perfect fluid matter field. Our
implementation is based on the Bondi metric, by which the spacetime is
foliated with a family of outgoing light cones. We use high-resolution
shock-capturing schemes to solve the fluid equations. The code can
accurately maintain long-term stability of a spherically symmetric,
relativistic, polytropic equilibrium model of a neutron star. In
axisymmetry, we demonstrate global energy conservation of a perturbed
neutron star in a compactified spacetime, for which the total energy
radiated away by gravitational waves corresponds to a significant
fraction of the Bondi mass.}

\section{Characteristic numerical relativity and hydrodynamics}

\subsection{Introduction}
We describe a new axisymmetric, fully general relativistic 
numerical code evolving the Einstein equations along with perfect 
fluid matter and apply it to studies of neutron stars.
Our numerical implementation of the field equations of general
relativity is based on the light cone formalism of Bondi\cite{BBM62}
and Tamburino-Winicour\cite{TaW66}. The light cone approach has a
number of advantages compared to spacelike foliations. i)~It is
physically motivated; the light cones offer a simple and unambiguous
physical gauge on which to base the numerical spacetime grid. ii)~It
is unconstrained; the evolved variables capture rather directly the
true degrees of freedom of the gravitational field. iii)~It is very
efficient; there are but two partial differential equations to solve,
along with a set of radial integrations. iv)~It allows for well
defined compactification of the domain, which leads to perfect outer
boundary conditions. v)~Finally, and may be most importantly, the
above theoretical advantages have been shown in a series of works to
translate to remarkably numerically robust and stable codes, see for
example~\cite{GLMW98}. For a recent review of the approach see
Winicour\cite{Win98}.

The broad target of this project is the detailed investigation of
neutron star dynamics using numerical relativity. A prerequisite for
such studies is the development of very accurate and long-term stable
general relativistic codes. It would seem that the feature list of the
characteristic approach makes it ideal for such studies. One serious
bottleneck of the approach immediately recognized by the trained
relativist is the breakdown of a lightlike coordinate system in the
emergence of light caustics. Interestingly though, due to its
quasispherical nature, no matter how agitated a neutron star, it is
unlikely to focus the light cones emanating from its interior.  A
limitation of more technical nature is the existence of a somewhat
restrictive Courant-Friedrichs-Lewy-condition for explicit algorithms
in multidimensions. Implicit evolution schemes would help in this
direction.

The incorporation of fluid matter fields in the characteristic
formulation of the Einstein equation was considered as early
as 1983\cite{IWW83} and numerical results were reported in reference\cite{BGLMW99}.  A
different line of attack was initiated recently\cite{PaF99} which
brings into the considerations the modern machinery from Computational
Fluid Mechanics. In this approach, the evolution equations for the
matter fields are solved using relativistic high-resolution
shock-capturing schemes\cite{Fon99,PaF99} (HRSC schemes) based upon
(exact or approximate) Riemann solvers. At this early stage of our
investigations, the ability of capturing shock waves in the
hydrodynamics is not essential. Nevertheless it will be so when
studying core-collapse scenarios. The implementation of HRSC schemes
in a general characteristic code without imposing symmetry conditions
is the current subject of a collaboration (GRACE, General relativistic
astrophysics via characteristic evolution). 

In this work we restrict ourselves to axisymmetric spacetimes. 
The code has been
developed building on previous work by G\'omez, Papadopoulos and
Winicour\cite{GPW94} (GPW), which constructed an axisymmetric
characteristic vacuum code, to which we have added a perfect fluid
matter field. Applications in spherical, dynamic black hole spacetimes
have been presented in references\cite{PaF99,PaF002}. Studies of
the spherically symmetric collapse of supermassive stars are discussed in
reference\cite{Lin00}.

\subsection{Fundamental equations}
\begin{figure}[h]
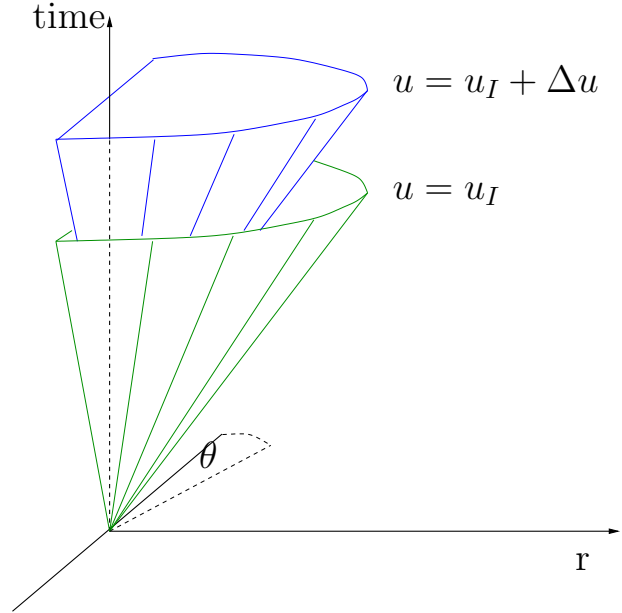

\begin{center}
        \input diagramme.pstex_t
\caption{Foliation of the spacetime by null hypersurfaces (light
cones): Starting from the geodesic of a freely falling particle, which
defines the origin of the coordinate system, surface forming light
rays are emitted.  We define a new time coordinate u to be constant
along each outgoing light cone.  The Killing coordinate $\phi$ is
suppressed in the diagram. The origin of the coordinate system will lie
(in the axisymmetric case) on the symmetry axis of the star.  If the
system has additional reflection symmetry with respect to the equator,
the origin is actually at the center of the star.  }
\label{fig1}
\end{center}
\end{figure}

The main equations we have to solve for the problem we are aiming at are
the Einstein equation
\begin{equation}
\label{Einstein}
G_{ab} - \kappa T_{ab} =  0
\end{equation}
and the conservation equation of the stress-energy tensor $T_{ab}$,
\begin{equation}
\label{fluid1}
\nabla^{a} T_{ab} = 0.
\end{equation}
The stress-energy tensor is chosen as that of a perfect fluid,
\begin{equation}
T_{ab} = \rho h u_{a} u_{b} + p g_{ab},
\end{equation}
where $\rho$ denotes the mass density, $h$ the relativistic specific
enthalpy and $u_{a}$ the four-velocity of the fluid. The pressure $p$
is subject to an equation of state. We choose a perfect fluid
equation of state
\begin{equation}
p = (\Gamma - 1) \rho \epsilon,
\end{equation}
where $\epsilon$ denotes the specific internal energy of the fluid, related to
the specific enthalpy as $h=1+\epsilon+\frac{p}{\rho}$.  In addition
we assume conservation of baryonic matter,
\begin{equation}
\label{fluid2}
\nabla^{a} (\rho u_{a}) =  0.
\end{equation}
We solve the above equations using the standard coordinate system of
characteristic numerical relativity (see Figure~\ref{fig1}).

More precisely, we use the Bondi metric
\begin{eqnarray}
\nonumber
ds^{2} & = & -(\frac{V}{r}e^{2 \beta} - U^{2} r^{2} e^{2 \gamma}) du^{2} - 2 e^{2 \beta} du \ dr 
- 2 U r^{2} e^{2 \gamma} du \ d\theta \\
\label{Bondi}
&  & + r^{2} (e^{2 \gamma} d
\theta^{2} + e^{-2 \gamma} sin^{2}\theta \ d \phi^{2}) 
\end{eqnarray}
with null coordinate $u$, radial coordinate $r$, azimuthal coordinate
$\theta$ and the spherical coordinate $\phi$, which is a Killing
coordinate. The four metric fields $\beta$, $\gamma$, $U$ and $V$ are
determined by solving the Einstein equation~(\ref{Einstein}). With a
choice of a coordinate chart $(x^{0},x^{i})$, the hydrodynamic
equations~(\ref{fluid1}) and (\ref{fluid2}) are written as an
explicit hyperbolic system, well-suited for numerical applications, as
detailed in references\cite{PaF99,PaF00}.

\subsection{Numerical implementation}

The numerical implementation of the Einstein equation~(\ref{Einstein})
closely follows that of GPW. Using the Bondi
metric~(\ref{Bondi}), equation (\ref{Einstein}) splits into
hypersurface equations on each light cone (for the fields $\beta$, $U$
and $V$) and one evolution equation (for $\gamma$), a wave equation.
These equations are solved by the same marching algorithms described
in GPW, now with additional source terms arising from the matter
fields. As we are using spherical coordinates, we have to take special
care with the numerical treatment of the coordinate singularities at
the origin and the polar axis. To this aim we extend the
treatment of the origin presented in GPW to nonvacuum spacetimes 
and regularize the
poles by introducing a new azimuthal coordinate $y=-cos\theta$ and by
redefining variables, for example $\gamma = \hat{\gamma} \ sin^{2}
\theta$.  As we want to keep the freedom of working with numerical
grids which only cover the neutron star without its vacuum exterior, we
generalize the radial coordinate used in GPW. Starting from an
equidistant radial coordinate $x \in [0,1]$, we allow for a general
coordinate transformation of the form $r = r(x)$, which enables us
to use compactified or non-compactified grids.

As already mentioned, we use HRSC schemes to solve the fluid
equations~(\ref{fluid1}) and (\ref{fluid2}). They constitute a
hyperbolic system of balance laws. We use a so-called method of lines
written in conservation form to solve these equations and to update the
initial data forward in time. The principal part is solved by a
Godunov-like method, with the same approximate Riemann solver as in
reference\cite{PaF99} (see this reference for further details).

\section{Tests and Applications}

In this section, we apply our coupled code to models of neutron
stars. For completeness we mention that the axisymmetric vacuum part
of the code has been successfully tested in additional situations. 
These comprise the comparison to an exact solution (SIMPLE), the
evolution of weak ingoing gravitational waves
and an energy conservation test for vacuum data, following the tests
described in~\cite{GPW94}.

\subsection{Long-term stability of spherical neutron stars}

As a simplified model for a self-gravitating neutron star we consider
the spherically symmetric solution of the general relativistic
hydrostatic equation with a polytropic equation of state, $p = K
\rho^{\Gamma}$. This is the Tolman-Oppenheimer-Volkoff solution in its
light cone representation~\cite{PaF99}. In our analysis, we have worked
with two representative models, following the choice of parameters
made also in~\cite{FSK99}. We restrict the discussion to the
more relativistic one of these models. In dimensionless units
($c=G=M_{\odot}=1$), the equilibrium properties of this star are
described by: a polytropic index $\Gamma=2$, a polytropic constant
$K = 100$ and a central density $\rho_{c}=1.28 \ 10^{-3}$. The
equilibrium model has a total mass $M = 1.4$. The time light needs to
travel across this neutron star corresponds to about 17u in our time
unit. We use this model for convergence tests as well as to check
long-term stability of the numerical algorithms.  

\begin{figure}[h]
\begin{center}
\includegraphics[angle=-90, width=0.8\textwidth]{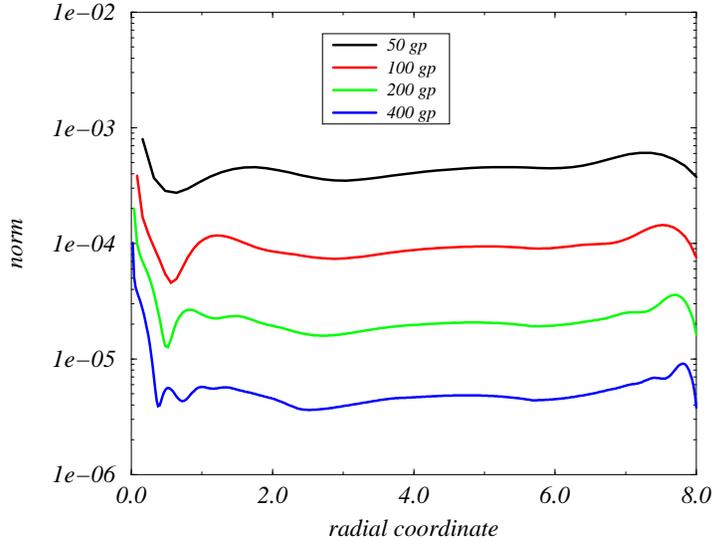}
\caption{Convergence test for the Tolmann-Oppenheimer-Volkoff
solution. Plotted is a suitable norm of all variables,
which measures the deviation from equilibrium, for the indicated
number of radial grid points (gp), after about
two light-crossing times.
To good approximation the solution is 2nd order convergent.
\label{fig2}}
\end{center}
\end{figure}

As the Tolman-Oppenheimer-Volkoff solution is static, convergence can
be easily checked by subtracting the evolved solution from the initial
(true) solution. We find second order convergence in all
variables. Figure~\ref{fig2} shows a suitable norm of all variables
as a function of the radial coordinate, which measures the deviation
from the initial solution for different grid resolutions.

\begin{figure}[p]
\begin{center}
\includegraphics[angle=-90, width=0.8\textwidth]{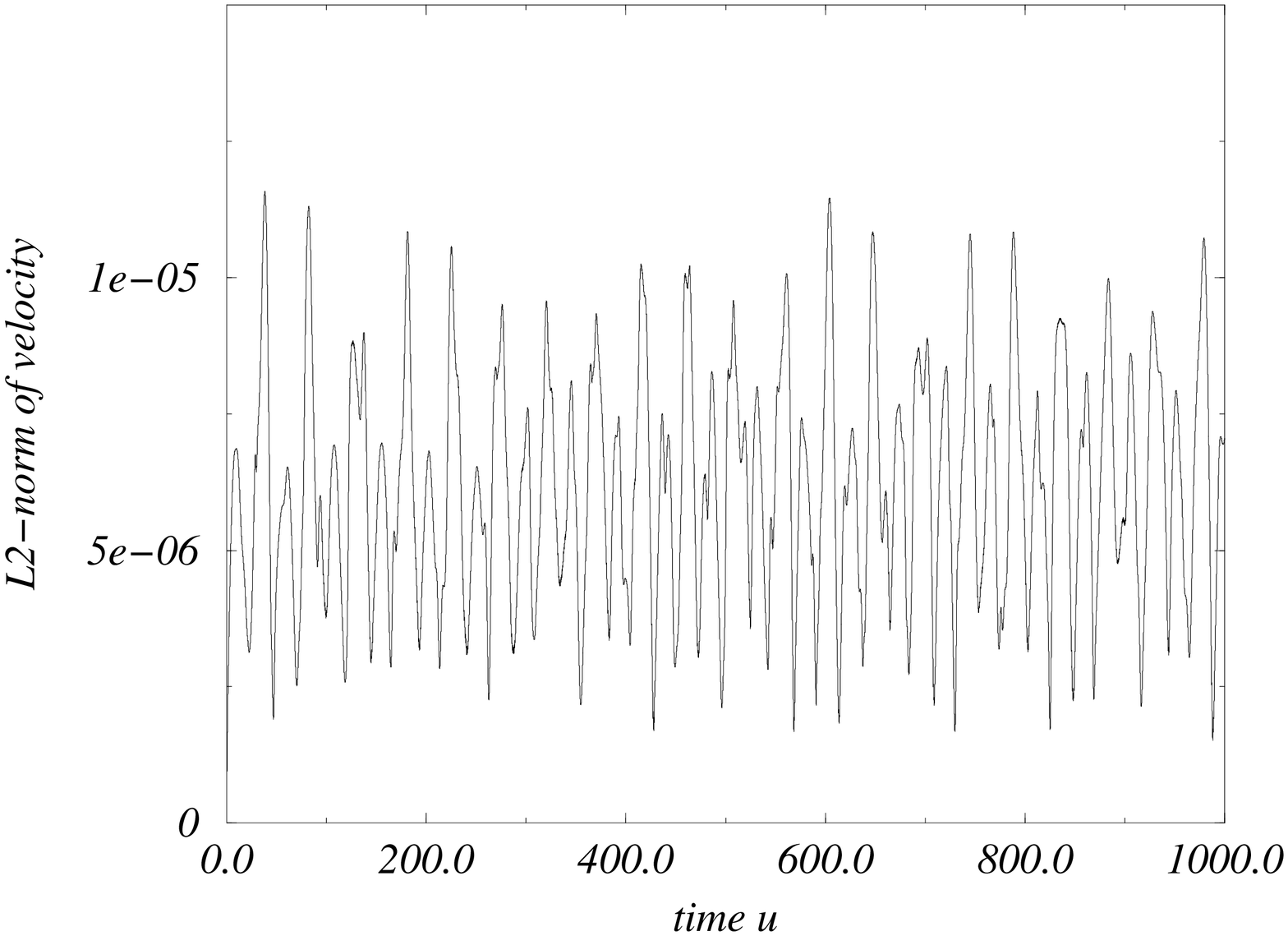}
\caption{Radial velocity of the star, averaged over the radial
coordinate, as a function of time. The evolution corresponds to
roughly $59$ light-crossing times.\label{fig3}}
\end{center}
\begin{center}
\includegraphics[angle=-90, width=0.8\textwidth]{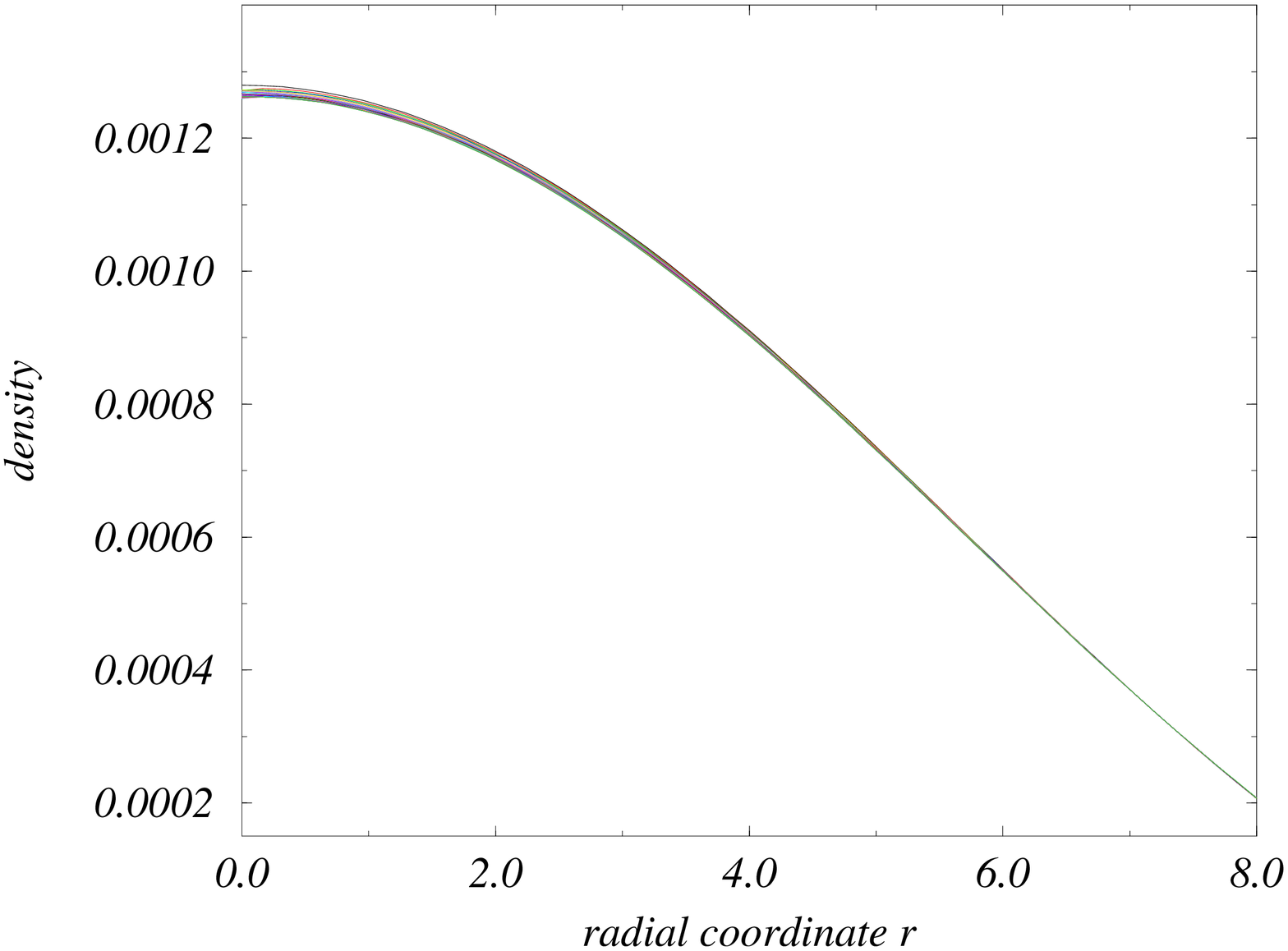}
\caption{Density profile of the neutron star at the initial and
subsequent times. The final time corresponds to $u=10000$ (roughly 590
light-crossing times). 
The equilibrium model is maintained to a very good precision after 
such very long integration times.
\label{fig4}}
\end{center}
\end{figure}

Figure~\ref{fig3} displays the L2-norm of the radial velocity
$||u^{r}||_{2} $ of the star
as a function of time. Due to discretization errors radial fluid
motion is generated, the radial velocity slightly deviates from zero
and the star oscillates in its radial modes of pulsation.  These radial
oscillations do not increase with time, which reflects the long-term
stability of our numerical implementation.

Figure~\ref{fig4} shows the density profile of the neutron star for a
very long integration time u=10000 for a lower resolution. 
Even though this corresponds to a very
long-term, fully general relativistic hydrodynamic evolution, the
density profile almost does not change, the equilibrium of the star
being maintained to very good precision. The upper line corresponds to
the initial model, the lowest line to the density profile at the time
u=10000, which corresponds to roughly 590 light-crossing times.

\subsection{Non-spherical matter motions}

\begin{figure}[h]
\noindent
\begin{minipage}[t]{.5\linewidth}
\center\epsfig{figure=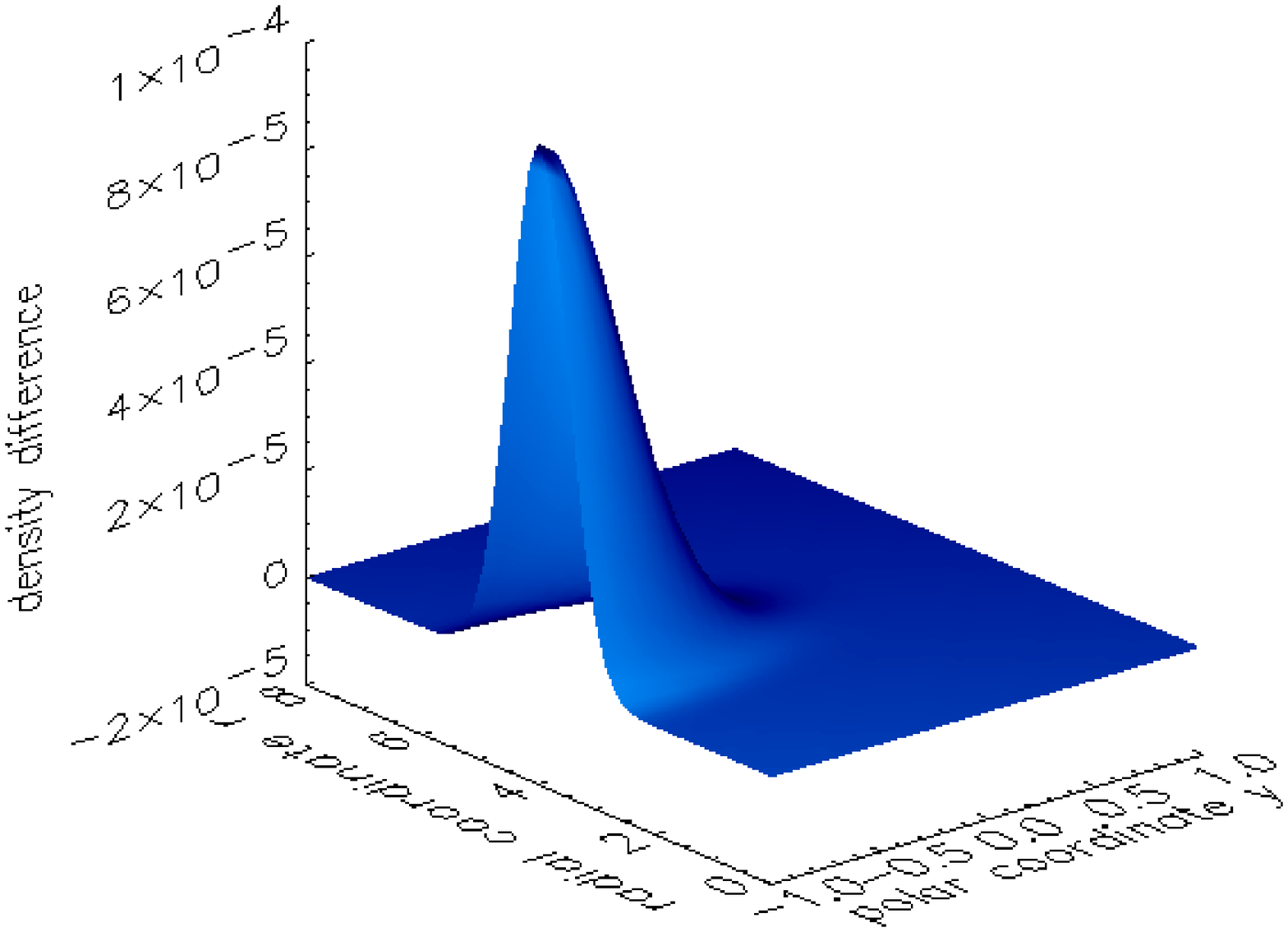,width=\linewidth}
u=0
\end{minipage}\hfill
\begin{minipage}[t]{.5\linewidth}
\center\epsfig{figure=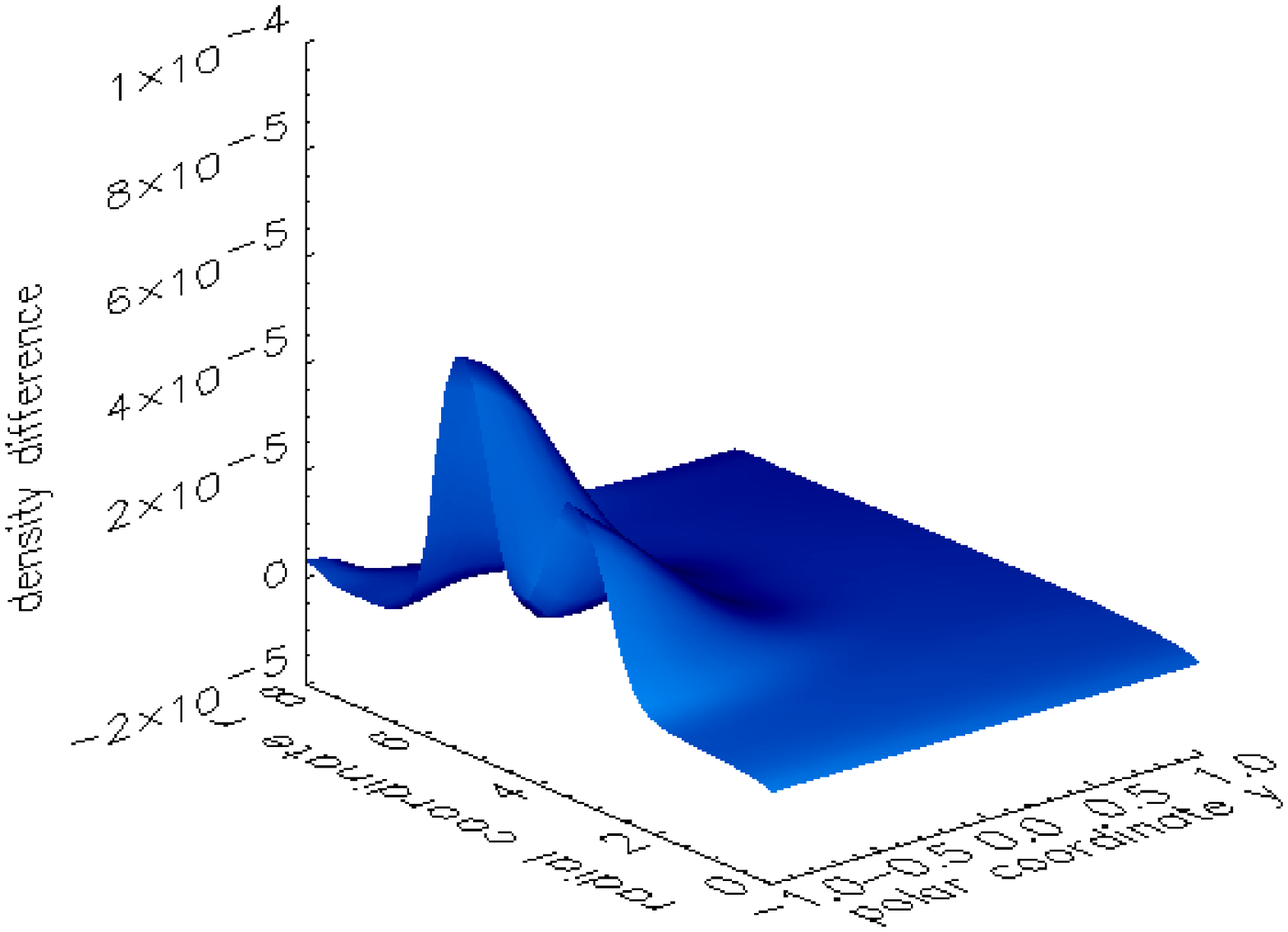,width=\linewidth}
u=2 
\end{minipage}
\noindent
\begin{minipage}[t]{.5\linewidth}
\center\epsfig{figure=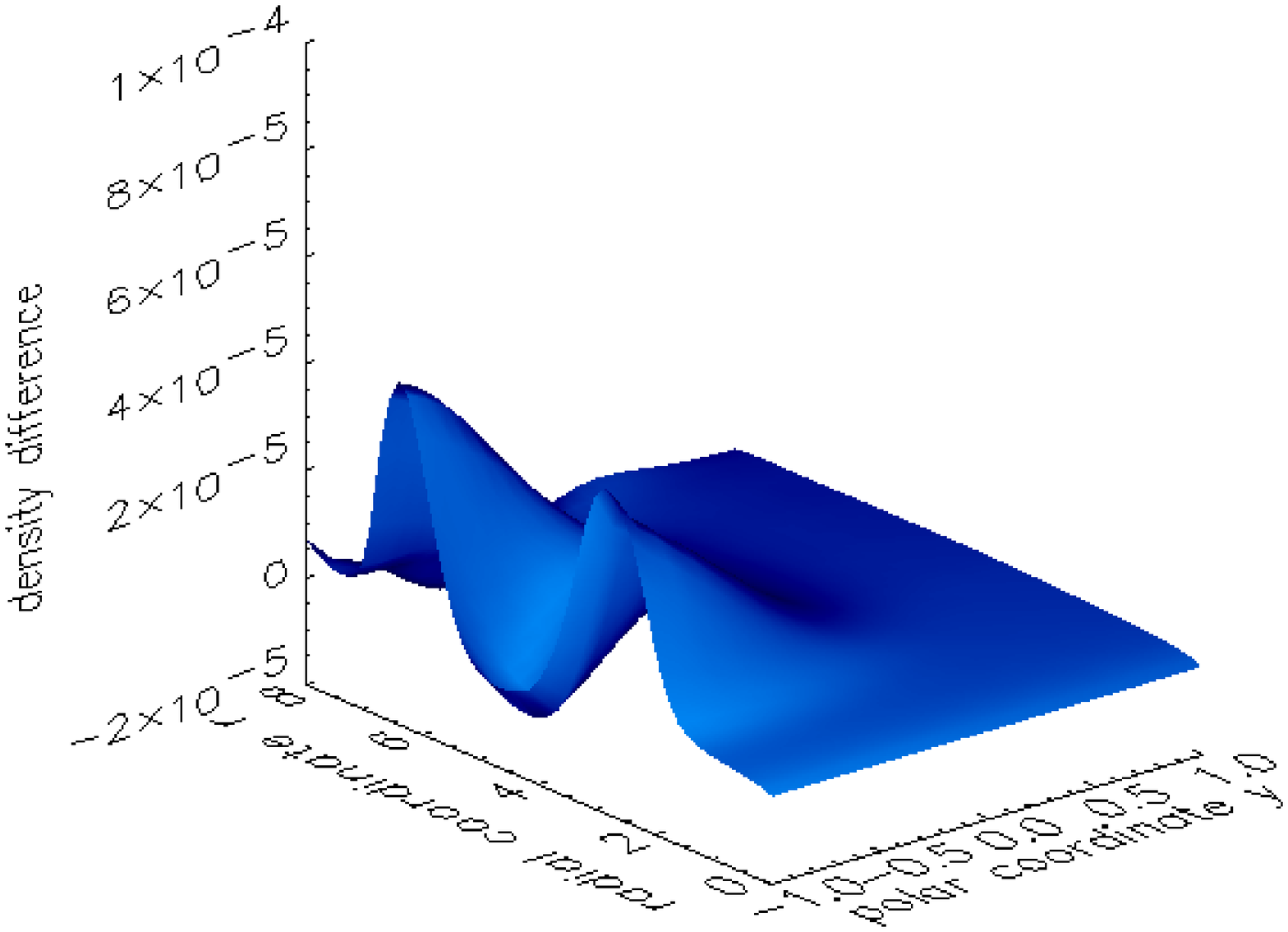,width=\linewidth}
u=4 
\end{minipage}\hfill
\begin{minipage}[t]{.5\linewidth}
\center\epsfig{figure=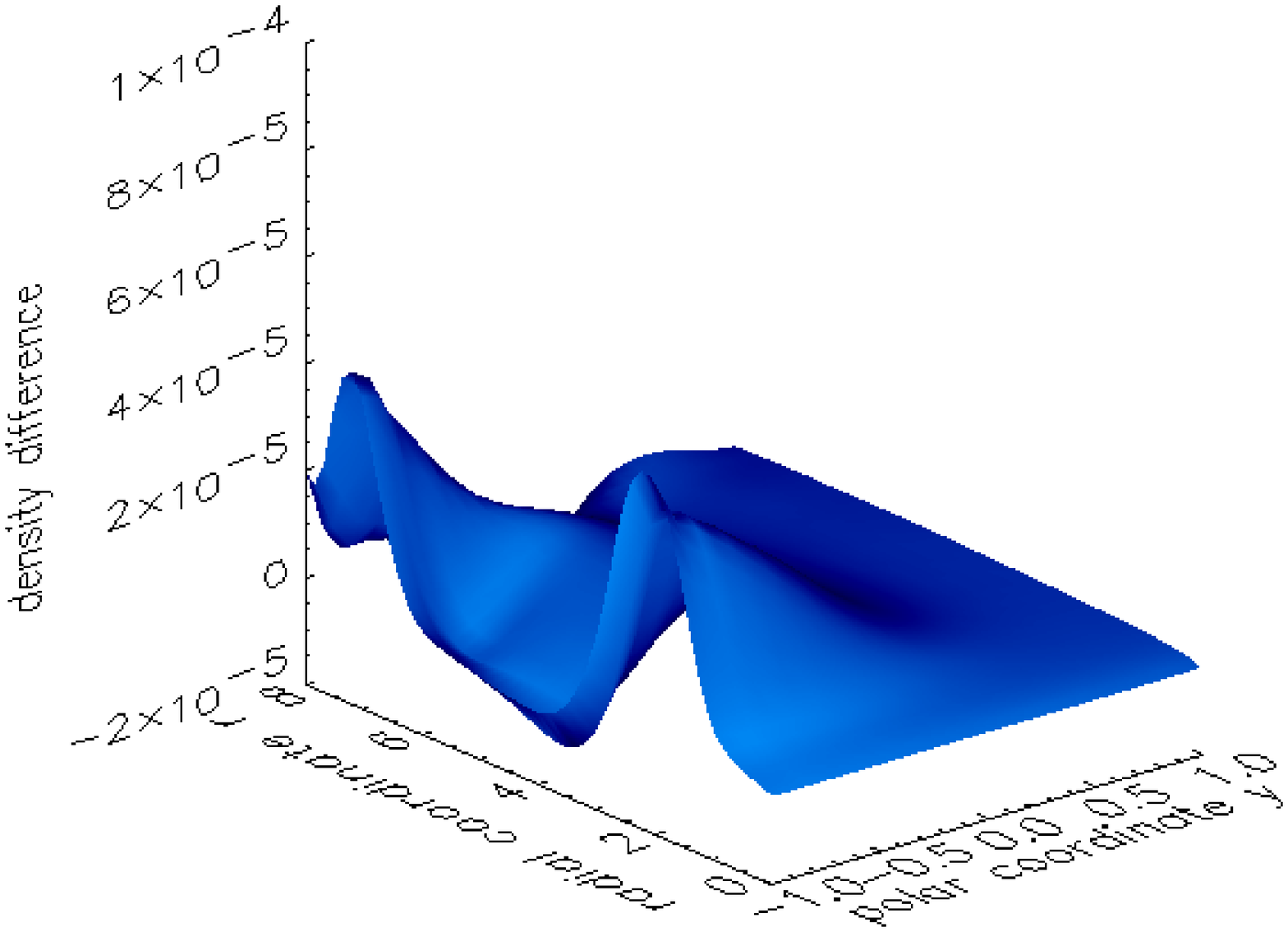,width=\linewidth}
u=6 
\end{minipage}
\vspace{0.5cm}
\caption{Evolution of a non-spherical density distribution on top 
of the equilibrium model of the neutron star. Plotted are the profile 
of the initial
overdensity and the profiles of the overdensity at subsequent
times as a function of the coordinates $r$ and $y$. 
The evolution shows a large, non-spherical mass transfer, which
allows us to check the second-order accurate implementation of all terms in our code.\label{fig5}}
\end{figure}

Next we consider the spherical model of the neutron star described in the
previous section and add a non-spherical density component.
This will lead to non-spherical
evolution and hence allows for testing the non-spherical terms of the
evolution system.  More precisely, we add a Gaussian profile, which is
centered at half radius at the polar axis and whose maximal value
corresponds to $10 \%$ of the density of the equilibrium model at that
point. Our initial model thus shows strong deviations from spherical
symmetry. The non-spherical fluid motion will trigger gravitational
waves.  Figure~\ref{fig5} shows a series of snapshots of this
evolution. We plot here the evolution of the overdensity, that is, the
difference between the non-spherical solution and the equilibrium one. The
evolution induces a large mass transfer. Due to this mass transfer, we
would expect a remarkable gravitational wave signal, i.e. the quantity
$\gamma$ picks up a non-zero contribution.  We found that our code
 is also second order convergent even for such
non-spherical data. As we are in a regime where the exact solution is
not known, we use Cauchy convergence for this test.

\subsection{Generation of gravitational waves}

One of the major benefits of using lightlike foliations is the
possibility of correctly extracting gravitational radiation after a
compactification of spacetime. For an asymptotically flat spacetime,
the metric fields can be Taylor expanded around future null infinity
$\cal{I}^{+}$ (which corresponds to the limit $r \to \infty$ using our
null foliation) in inverse powers of the radial coordinate $r$. Using
the coefficients of this power series, the radiated energy emitted by
gravitational waves can be determined.\cite{IWW83} In the same way,
one could determine the Bondi mass, the total energy. However, as one
would have to pick up nonleading terms in the power series expansion for the
metric variable $V$, as the leading terms diverge on $\cal{I}^{+}$,
the numerical extraction is easily spoiled by numerical errors. We
therefore follow the work of G\'omez et al\cite{GRWI93} and globally
introduce new metric variables, from which the Bondi mass can be
determined using the leading term of their power series expansion.

\begin{figure}
\begin{center}
\includegraphics[angle=-90, width=0.8\textwidth]{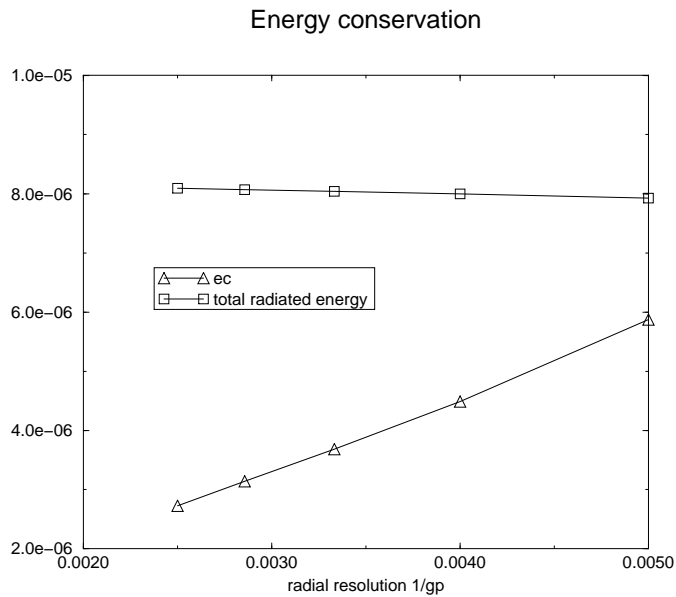}
\caption{Global energy conservation test for a neutron star and a
strong gravitational wave. Plotted are the deviation from energy
conservation (triangles) and the total energy emitted by
gravitational waves (squares). The final integration time is $u_{*}=0.002$,
the number of angular grid points is 0.2 gp + 1, where gp denotes the number
of radial grid points. \label{fig6}}
\end{center}
\end{figure}

In this section, we focus on a global energy conservation test. We
start with a strongly perturbed neutron star and evolve this
configuration for a very short time, which is sufficient for our purpose
of testing convergence. We use a strong ingoing gravitational wave to
perturb the equilibrium star,
\begin{equation}
\hat{\gamma}= 0.05 \ e^{-2 (r-4)^{2}} \ e^{-4 y^{2}}.
\end{equation}
Such a large amplitude is not realistic, but we choose it to test our
numerical implementation in the nonlinear regime. Such initial data
involve large velocities in the fluid of the star and produce strong 
outgoing gravitational
waves. By 'strong' we understand that the energy which is
radiated away by gravitational waves is larger than the numerical
errors for the calculation of the Bondi mass for the chosen resolution
and integration time.  Let $M$ be the Bondi mass and $P$ the total energy
radiated away by gravitational waves. Thus, the convergence of the quantity
\begin{equation}
ec := M|_{u=0} - M|_{u=u_{*}>0} - P|_{[0,u_{*}]}
\end{equation} 
to zero represents a very severe global test for our implementation (see
Figure~\ref{fig6}).

The obtained first-order convergence rate can be explained by the use
of a {\it total variation diminishing} HRSC scheme, which, although
it is second-order accurate in smooth, monotonous parts of the flow,
reduces to first-order at local extrema, which are present in the
interior of the numerical domain in this test. Tests including
propagation and scattering off the origin of pure vacuum
gravitational fields yielded the expected second-order convergence.

There are two additional consistency conditions, which relate the
metric quantities on $\cal{I}^{+}$.\cite{Pap93} These are
\begin{eqnarray}
S & =  & U_{,\theta} + U \ cot\theta \\
\gamma_{,u} & = & -\frac{1}{2}e^{-2 \gamma} sin\theta(e^{2 \gamma}
\frac{U}{sin \theta})_{,\theta},
\end{eqnarray}
where $S$ is globally defined as
\begin{equation}
S = \frac{V-r}{r^{2}}.
\end{equation}
Here again, we find first-order convergence.

\section{Conclusion}
We have developed an axisymmetric, fully general relativistic
characteristic code with a perfect fluid matter field. We have checked
convergence in spacetimes containing a neutron star.  For a spherical,
relativistic equilibrium model of a polytropic neutron star, we have
shown long-term stability for hundreds of
light-crossing times. In axisymmetry, we analyzed convergence in
strongly perturbed neutron stars. Compactifying the spacetime, we
could show global energy conservation, in the sense that the amount of
energy lost by the system is radiated away by gravitational waves.

\section*{Acknowledgments}
We thank Felix Linke for helpful discussions.  F.S. thanks the
Relativity and Cosmology Group at the University of Portsmouth for
hospitality. P.P. warmly thanks the MPI for the hospitality during the
Oktoberfest, which helped this project get underway.

\end{document}

%% file: diagramme.pstex_t
\begin{picture}(0,0)%
\epsfig{file=diagramme.pstex}%
\end{picture}%
\setlength{\unitlength}{1776sp}%
\begingroup\makeatletter\ifx\SetFigFont\undefined%
\gdef\SetFigFont#1#2#3#4#5{%
  \reset@font\fontsize{#1}{#2pt}%
  \fontfamily{#3}\fontseries{#4}\fontshape{#5}%
  \selectfont}%
\fi\endgroup%
\begin{picture}(8525,8490)(1023,-8701)
\put(1351,-511){\makebox(0,0)[lb]{\smash{\SetFigFont{14}{16.8}{\familydefault}{\mddefault}{\updefault}time}}}
\put(6376,-1411){\makebox(0,0)[lb]{\smash{\SetFigFont{14}{16.8}{\familydefault}{\mddefault}{\updefault}$u = u_{I}+\Delta u$}}}
\put(6376,-2911){\makebox(0,0)[lb]{\smash{\SetFigFont{14}{16.8}{\familydefault}{\mddefault}{\updefault}$u = u_{I}$}}}
\put(3676,-6661){\makebox(0,0)[lb]{\smash{\SetFigFont{14}{16.8}{\familydefault}{\mddefault}{\updefault}$\theta$}}}
\put(8926,-8086){\makebox(0,0)[lb]{\smash{\SetFigFont{14}{16.8}{\familydefault}{\mddefault}{\updefault}r}}}
\end{picture}